\documentclass[prl,aps,twocolumn,superscriptaddress,preprintnumbers,amsmath,amssymb]{revtex4}
\usepackage{graphicx}
\usepackage{dcolumn}
\usepackage{bm}
\usepackage{longtable}
\usepackage{xcolor}
\usepackage[normalem]{ulem}
\begin{document}
\title{Spin orientation - a subtle interplay between strain and multipole Coulomb interactions}
\author{Subhra Sen Gupta}
\email{subhra.sengupta@snu.edu.in}
\affiliation{Department of Physics, Shiv Nadar Institution of Eminence (SNIoE), Gautam Buddha Nagar, Uttar Pradesh 201314, India.}
\affiliation{Department of Condensed Matter and  Materials Physics, S.N. Bose National
Center for Basic Sciences, Kolkata - 700106, India.}
\author{Shinjini Paul}
\affiliation{Department of Condensed Matter and  Materials Physics, S.N. Bose National
Center for Basic Sciences, Kolkata - 700106, India.}
\author{Suman Mandal}
\affiliation{Surface Physics Division, Saha Institute of Nuclear Physics, Kolkata - 700064, India.}
\author{D. D. Sarma}
\affiliation{Solid State and Structural Chemistry Unit,
Indian Institute of Science, Bangalore - 560012, India.}
\author{Priya Mahadevan}
\email{priya@bose.res.in}
\affiliation{Department of Condensed Matter and  Materials Physics, S.N. Bose National
Center for Basic Sciences, Kolkata - 700106, India.}
\email{priya@bose.res.in}
\date{\today}

\begin{abstract}
We address the technologically important issue of the spin orientation
on a correlated magnetic surface and how to manipulate it. We consider
a prototypical strongly correlated system, NiO, and show that a single particle approach with anisotropic hoppings,
or even a many-electron model with a scalar Hubbard $U$ and Hund's $J$ fails to explain the strain
driven spin reorientation transition (SRT). We set up a model treating both anisotropic single particle effects
and orbital-dependent, full multipole electron-electron interaction effects at the same footing. Within this model,
predictive power to explain the observed SRT is regained and the results indicate the
novel possibility of using an electric field to control SRT in magnetic films grown on piezoelectric
substrates.
\end{abstract}










\maketitle

Recent advances in growth technology have made it possible to grow
thin films of transition metal oxides with the same quality that was
earlier possible only for semiconductors \cite{oxide1, oxide2}. Technological
considerations have driven the study of such materials, with the
aim of exploiting the magnetic properties or in some cases the
magnetoelectric properties in devices constructed. A
key property of a ferromagnet is the direction of its magnetization,
which should be fixed rigidly in permanent magnets. However in
sensing applications as in read-heads, one requires the
magnetization to be easily rotated by a small applied field \cite{mae1,mae12,mae13}. As
this property of the material is governed by the magneto-crystalline
anisotropy (MCA), one requires a microscopic understanding of the
phenomenon, to be able to find the optimal material for any specific
application \cite{mae2,mae22,mae23,mae24}. In recent times there has been a lot of interest 
in materials with a {\em perpendicular magnetocrystalline anisotropy} (PMA) for storage applications.

Apart from multilayers of metals which have been a staple ingredient of GMR devices, 
transition metal oxide heterostructures have been widely studied
with unusual magnetoelectric phenomena being predicted.
 Growing materials on different substrates and thereby straining them, has been found to be a way to engineer the orientation of the spin \cite{lin_a,Dashwood,Wang}. Apart from strain effects, dipolar effects are often invoked in
discussions of spin orientation transitions involving assemblies of particles,
while shape anisotropy is often invoked to discuss these in particles with different shapes.
Even in the same system, the behavior may vary as a function of thickness\cite{Oepen,Slezak1, Slezak2,Allenspach},
leading to the origin of the different spin orientations being attributed to surface anisotropy when the thickness 
is small, while bulk anisotropy effects take over for thicker samples.
In this work we examine in detail the example where the spin reorientation
transitions have been  achieved by controlling the lattice strain. It is surprising that despite the fact that the topic of spin reorientation has been studied for several decades, there is no consensus on the mechanism.  

Considering the example of NiO which is like the fruit fly of correlated electron physics, one finds that films grown on MgO, presenting a
tensile stress, are found to favor an out-of-plane spin alignment 
\cite{NiO-MgO-expt}, while
those grown on Ag with a smaller lattice constant, implying compressive
strain \cite{NiO-Ag-expt}, are found to favor an
in-plane spin alignment. Interestingly, the microscopic origin of such a
SRT as a function of  strain is still
unclear. NiO is expected to have a quenched orbital moment, on the
basis of orbital filling though there is experimental evidence
\cite{orb-mom} of a partial revival of the orbital moment.
This alternate route using strain to control SRTs suggests that
a piezoelectric substrate would allow such transitions to take
place on a single substrate by tuning the electric field or more recently the 
growth of free standing membranes of various oxides has allowed strain to become
a continuous parameter \cite{Xu,Lu,Peng,Cai,Dong2,Jin}, thereby opening up new possibilities for all such magnetic devices.

We have carried out electronic structure calculations within
a plane wave implementation of density functional theory with projected augmented potentials \cite{Blochl, Kresse} 
in the Vienna Ab-initio Simulation Package (VASP) \cite{2Kresse,3Kresse,4Kresse,5Kresse}. The in-plane lattice parameters were fixed 
to that of the substrate \cite{NiO-MgO-expt, NiO-Ag-expt}, while the out of plane lattice parameter was optimised within the calculations. 
This led to the lattice parameters 
a=4.212 $\AA$ and c=4.110 $\AA$ for the MgO substrate and a=4.092$\AA$ and c=4.248 $\AA$ for the Ag substrate.  The electronic
structure was determined using a cut off energy of 800 eV for the kinetic energy of the plane waves included in the basis, in
addition to a k-points mesh of 16$\times$16$\times$16 for the integrations in momentum space. The generalised gradient approximation (GGA) \cite{Perdew}
was used for the exchange-correlation functional, with the effect of electron-electron interactions included at the GGA+U level \cite{Dudarev}.
The antiferromagnetic structure of unstrained NiO was assumed in these calculations and the preferred orientation of the spin
was determined. The results were cross-checked with  a full potential linearized augmented plane
wave implementation of density functional theory \cite{wien2k} and were found to be consistent. Further details are found in the Supplementary Information. These calculations could reproduce the experimentally observed SRT.
\newline
In order to understand the origin, we considered a representative many body Hamiltonian 
for both strained cases. 
Model Hamiltonian calculations were carried out for appropriately strained 
NiO$_{6}$ clusters, representing the local environment of Ni present in the thin films. 
The structural information of the NiO$_6$ octahedra have been obtained from ab-initio calculations.
The orbital dependent many-body Coulomb interactions have been treated exactly along with
the single-particle part of the Hamiltonian.  The calculations are found to
reproduce several experimental trends. An important result that emerges
from our calculations is that both anisotropic hoppings as well as a scalar Coulomb interaction 
are insufficient to capture the phenomenon. The crux of the matter is that
a spin and orbitally differentiated Coulomb interaction results in 
excited states that are {\it qualitatively} different from
that expected from a scalar ($U$,$J$). These are then mixed into the 
orbitally non-degenerate ground state via spin-orbit interactions resulting in the observed SRT.

On the basis of the many-body cluster calculations,
we first examine the case of NiO on MgO where the octahedron is {\em expanded
in-plane}. The values of the various moments 
within the many-body calculations are
$\langle S_{z}\rangle =-0.92$,
and $\langle S \rangle = 0.94$, a large $z$-component of the spin,
nearly equal to $\langle S \rangle$ is found. This implies that the spin
points out of plane. Moreover, $ \langle L_z \rangle $=-0.22 and 
$\langle L \rangle$=2.87, established that there is a large, but
incomplete quenching of orbital moment. 
In sharp contrast, the moments are $\langle L_{z} \rangle
\sim 0$ , $\langle S_{z}\rangle \sim 0$ , $\langle L
\rangle = 2.86$, $\langle S \rangle = 0.93$, 
for the contracted in-plane case  of NiO on Ag, 
indicating strong changes
compared to the previous case. Clearly, while $\langle L
\rangle$ and $\langle S \rangle$ are almost identical in the two
cases, $\langle L_{z} \rangle$ and  $\langle S_{z}\rangle$ are
nearly zero only for the case of in-plane contraction. 
This implies that the spin lies in-plane in this case as
observed experimentally, along with a complete quenching
of the orbital moment. Thus, the model Hamiltonian approach 
adopted here is able to reproduce the observed spin reorientation in 
NiO as a function of the compressive/tensile
strain of the NiO overlayer. 
It is important to note here that two calculations, representing NiO
on MgO~\cite{NiO-MgO-expt} and Ag~\cite{NiO-Ag-expt} have essentially identical
parameter sets except for the sole difference of scaling the
hopping energies by their dependence on the distance between 
the two centres (Ni and O).Thus, the observed SRT are in absence of any adjustable or tuning
of parameter strengths, giving confidence to these results.

To understand the microscopic origin of the SRT, it is instructive to carry out a 
second-order perturbative analysis in the {\em spin-orbit} (SO) interaction. 
~\cite{balhausen,yoshida} 
The ground state has a quenched orbital angular momentum.
Treating the spin-orbit interaction to second-order in perturbation theory  
yields the spin Hamiltonian :
\begin{equation}
{\cal H}_{aniso} = DS_{z}^{2} + E(S_{x}^{2} - S_{y}^{2}) \label{spin-hamil}
\end{equation}
where, the first term 
represents  {\em perpendicular anisotropy}, with
$$D = \lambda^{2}\left[ \frac{1}{2}(\Lambda_{x} + \Lambda_{y}) - \Lambda_{z} \right]$$ and the second term
yields an {\em
in-plane anisotropy} with
$$E = \frac{\lambda^{2}}{2}(\Lambda_{y} - \Lambda_{x})$$
In the above expressions the $\Lambda_{i}$ ($i=x,~y,~z$) 
are principal
axis values of the tensor $\Lambda_{\mu\nu}$ given by,
\begin{equation}
\Lambda_{\mu\nu} = \sum_{n} \frac{\langle 0|L_{\mu}|n\rangle \langle n|L_{\nu}|0\rangle }{E_{n}-E_{0}},
~~(\mu,\nu = x,y,z) \label{lambda-tensor}
\end{equation}
$E_{0}$, $E_{n}$ are the ground and excited state many-body energies of the {\em unperturbed}
microscopic Hamiltonian, while $|0\rangle$ and $|n\rangle$ are the
corresponding eigenvectors. L$_{\mu}$ ($\mu=x,y,z$) are the components of the orbital angular momentum operator.

The unperturbed Hamiltonian ($\lambda=0$) is solved for the {\em contracted} and the {\em expanded} cases by exact diagonalization. The finite matrix elements of the $L_{x,y,z}$ operator between the excited states and the orbitally quenched ground state, are listed as a table in the {\em SM}. Interestingly, we find that only the first two crystal-field excited states determine the sign and magnitude of the anisotropy constants. While one would normally expect this to be an {\em energy denominator effect}, we find that it is largely driven here by symmetry as a result of which most of the higher matrix elements of $L_{\mu}$ 
turn out to be zero. The detailed symmetry arguments, is discussed in detail in the {\em SM}.

As a result of the inherent in-plane symmetry of the 
tetragonally distorted NiO, we have the diagonal members
$\Lambda_{x}=\Lambda_{y} \ne \Lambda_{z}$ and the off-diagonal
elements are zero. Thus, the expressions for $D$ and $E$ simplify to
: $D = \lambda^{2}(\Lambda_{x} - \Lambda_{z})$ and $E=0$. 
Thus, Eqn.~(\ref{spin-hamil}) reduces to $H_{aniso}= D S^2_z$. From our exact diagonalization calculations for the NiO$_{6}$ cluster,
we find $D =-2.23 \times 10^{-4}$~eV $< 0$ for the {\em expanded in-plane} 
case. This implies that $S_{z}$ would try to maximize
itself in magnitude, in order to minimize the
energy. This is indeed seen from the full (non-perturbative) calculation 
as also from experiments. This gives rise to a moment pointing almost completely
out-of-plane with $\langle S_z \rangle \sim \langle S \rangle$. 
This is distinct from the usual shape anisotropy of dipolar origin. 
For the NiO$_6$ octahedron {\em contracted in-plane}
we find $D = +2.28 \times 10^{-4}$~eV $> 0$.
This implies that $S_{z}$ would try to set
itself to zero to minimize the energy, just as
we find from the full calculation. This is the origin of the spin-flop
state with vanishing out-of-plane moment,~\cite{NiO-Ag-expt} retaining
the same large value of $\langle S \rangle$.

\begin{figure}
\begin{center}
\includegraphics[angle=0, width=0.5\textwidth]{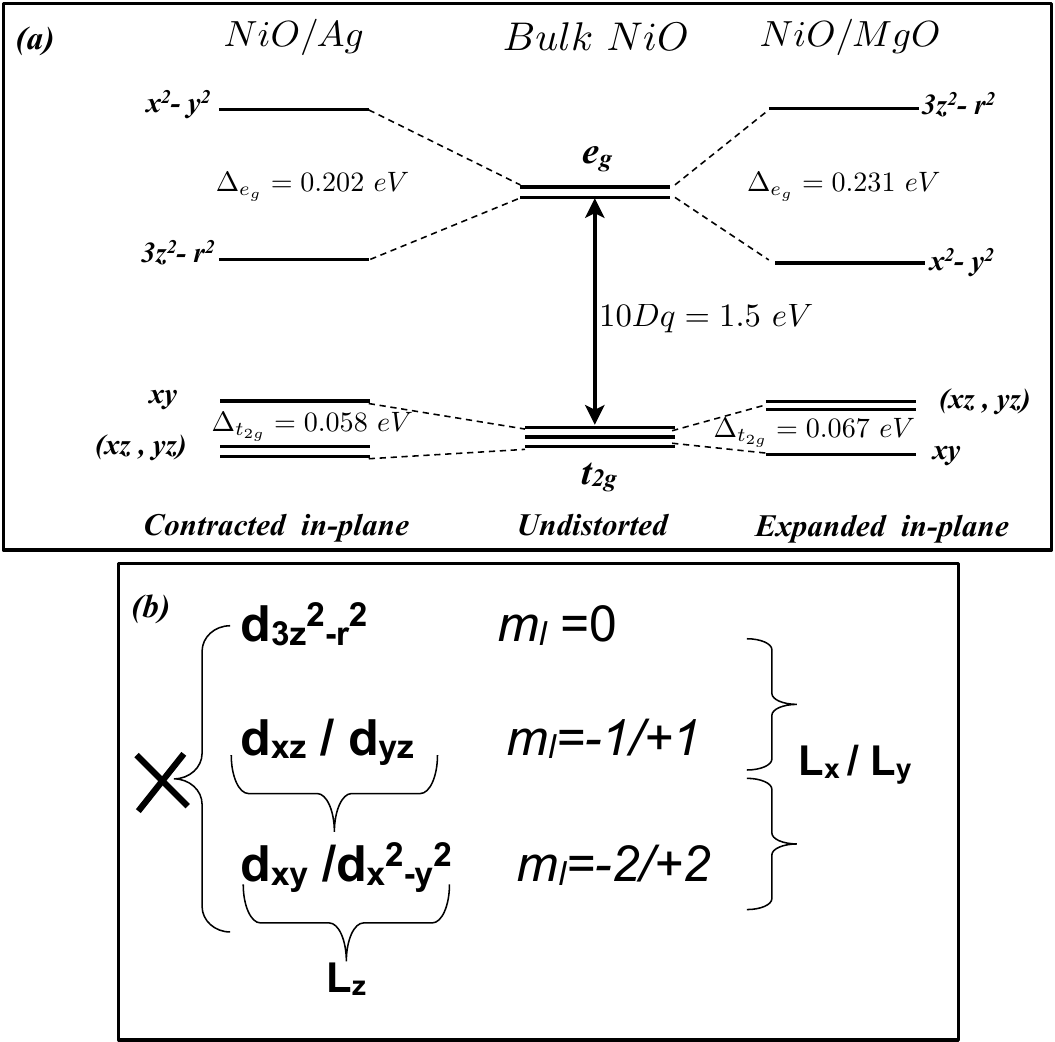}
\caption {(a) Schematic showing the nature and magnitude of crystal field splitting of the $3d$-levels on the Ni atom for bulk (undistorted) NiO (centre), and NiO thin-films grown on Ag (left) and on MgO (right) respectively (both tetragonally distorted). (b) Listing of the various {\em real} $d$ orbitals in terms of its $m_{l}$ basis components, and also a listing of how these states are connected by the components of the $\vec{L}$ operator.} \label{fig1}
\end{center}
\end{figure}

\begin{figure}
\begin{center}
\vspace{-1.0cm}
\includegraphics[angle=0, width=0.5\textwidth]{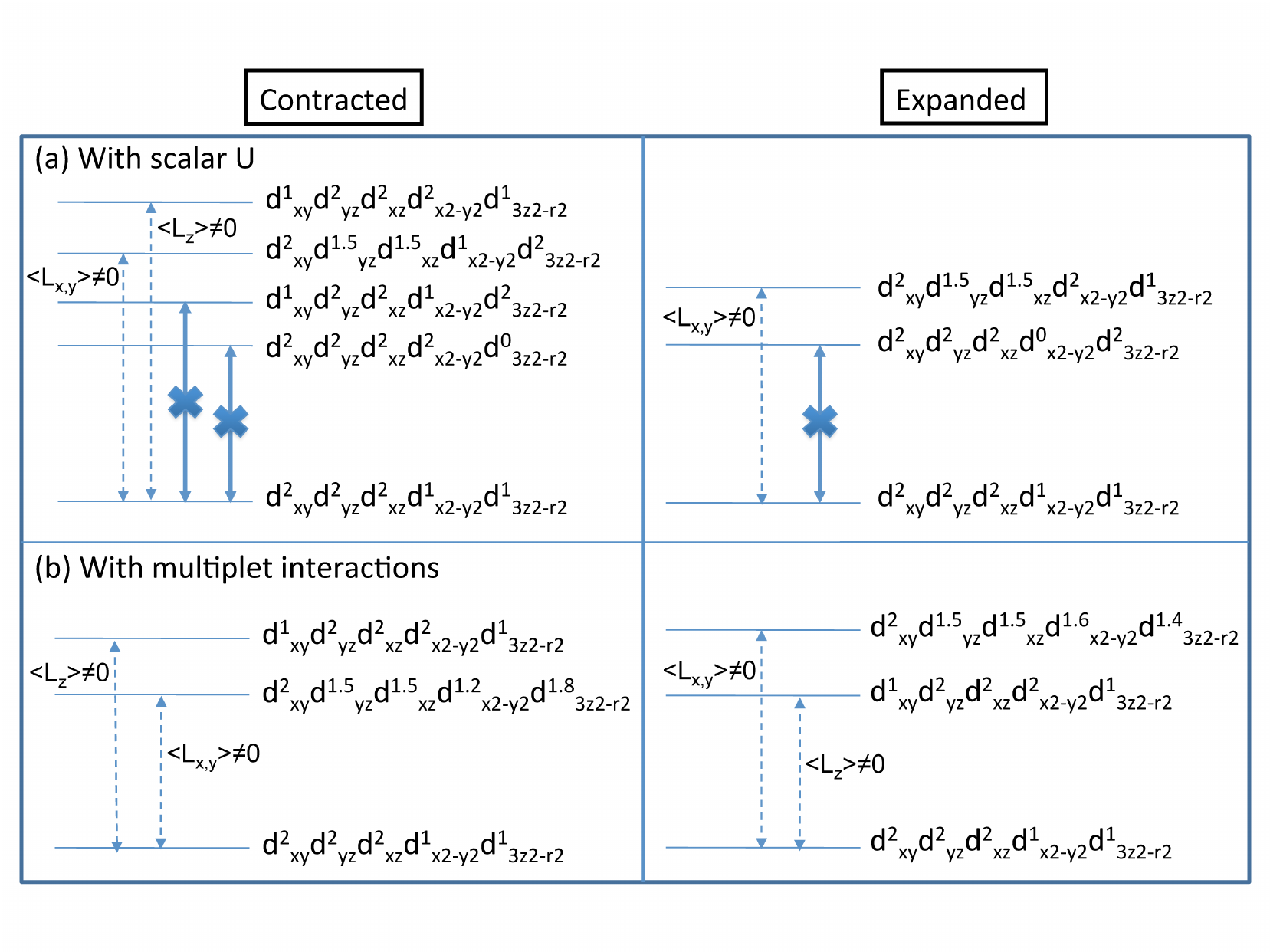}
\vspace{-0.8cm}
\caption {(\emph{color online})  Schematic representation of the many-body ground state and relevant excited states, in terms of single-particle orbital occupancies, for the contracted in-plane case of NiO/Ag ({\em left}) and for the expanded in-plane case of NiO/MgO ({\em right}). {\em Row} (a) lists this for the case of a {\em scalar} $U$, while {\em row} (b) lists this for a more realistic case with multipole Coulomb interactions (multiplets). It also shows which matrix elements $\langle n|L_{\mu}|0\rangle$ are finite in each case.} \label{fig2}
\end{center}
\end{figure}

Thus, the sign of $D$, which determines the orientation of the Ni spin, is governed by the relative strengths
of $\Lambda_{x}$ and $\Lambda_{z}$, since $D = \lambda^2 (\Lambda_{x} - \Lambda_{z})$. We analyze 
below the reasons for the change in the relative strengths of $\Lambda_{x}$ and $\Lambda_{z}$, 
when the epitaxial strain is changed from {\em tensile} to {\em compressive}. 
An analysis of the occupancies of the orbitals for both the ground state as well as excited states 
reveals the important role of orbital-dependent Coulomb interactions.
The most obvious change of the
electronic structure between the expanded and contracted cases 
at the single particle level is reflected
in the ordering of the electronic orbital energies.
Independent of the nature of the distortion of the NiO$_6$ octahedron,
the exchange splitting between the up- and the down-spin electrons is
much larger than the crystal-field splitting between the $t_{2g}$
and the $e_{g}$ groups of orbitals, for these narrow band systems, as also confirmed by {\it ab-initio} results. Thus, the down-spin $t_{2g}$
and $e_g$ states are filled by five electrons, independent of the 
nature and the extent of the distortion, the remaining three electrons
occupy the $t_{2g}$ up-spin orbitals, leaving the $e_g$ up-spin
orbitals empty. We first examine the case in which the octahedron is {\em contracted
in-plane}.  Here the $d$
orbitals with $t_{2g}$ symmetry are split into lower-lying doubly degenerate $d_{xz}$
and $d_{yz}$ states, and a higher-lying $d_{xy}$ state, while the
$d$ states with $e_g$ symmetry are split into a
lower $d_{3z^2-r^2}$ and a higher
$d_{x^2-y^2}$ orbital, as shown in Fig.~\ref{fig1}(a) ({\em left panel}). The corresponding energy levels for undistorted bulk NiO is shown in the {\em central panel}, where the $t_{2g}$ and $e_{g}$ levels are not split any further.
The energy level orderings are reversed for the expanded in-plane, as
shown in Fig.~\ref{fig1}(a) ({\em right panel}). For the 2$\%$ contracted in-plane case, in addition to an octahedral splitting of
$10Dq\sim1.5$ eV (1 eV arises from bare crystal field effects while 0.5~eV arises
from hybridization with the ligand) between
$t_{2g}$ and  $e_{g}$ levels, the tetragonal distortion induces the splittings, $\Delta_{e_{g}}$ $\sim$
0.202 eV and $\Delta_{t_{2g}}$ $\sim$ 0.058 eV, for the $e_g$ and $t_{2g}$ levels respectively, as shown in Fig.~\ref{fig1}(a).

It turns out that both the ground state, as well as the excited eigenstates originating from the $^{3}$F, which are not significantly modified by CI with those originating from $^{3}$P, have diagonal density matrices $\rho_{\alpha\beta}=\langle n|a^{\dagger}_{\alpha}a_{\beta}|n\rangle$ ($n$=excited state index) in the {\em real basis} $\alpha,\beta=\{d_{xy},d_{yz},d_{xz},d_{x^{2}-y^{2},d_{3z^{2}-r^{2}}}\}$. Hence these excited states can be conceptually expressed as single particle transitions over the ground state.
The ground state occupancies have already been discussed earlier. 
The excited states in presence of orbitally resolved, multipole Coulomb interactions are qualitatively different from what is intuitively expected in a single-particle picture or that with a scalar $U$ and $J$. The {\em first excited state} is found to be a {\em spin-preserving} excitation from
the lower-lying doubly degenerate $(d_{xz}$, $d_{yz})$ level, primarily into the $d_{3z^2-r^2}$ orbital (80\%)
with some admixture of $d_{x^2-y^2}$ (20\%) as indicated in Fig.~\ref{fig2}(b). 
The {\em second excited state} is found to result from a {\em spin-preserving} transition from the higher 
$t_{2g}$ orbital ($d_{xy}$) to the higher unoccupied $e_{g}$ orbital ($d_{x^2-y^2}$) (Fig.~\ref{fig2}(b)). Fig.~\ref{fig1}(b) lists the $m_{l}$ basis states which contribute to each of the real basis $d$ states, and also lists the matrix elements of $L_{\mu}$ ($\mu=x,y,z$) between these states. From the above, it is clear that the ground state is connected to two states in the {\em first crystal-field excited manifold} via $L_{x,y}$, and not $L_{z}$, while the situation is reversed for the {\em second crystal-field excited manifold} one state of which is connected to the ground state via $L_{z}$ alone. This is confirmed by the results of actual calculations of matrix elements as detailed in the {\em SM}. These first two excited levels are the key to determining the sign of $D$, as the other finite matrix elements are much smaller in comparison.
We find that the matrix elements, aided by the increasing energy denominators as one goes from the first to the second excited state, dictates that $\Lambda_{x}>\Lambda_{z}$ which yields $D>0$.

For the case of NiO on MgO, where the NiO$_{6}$ octahedron is expanded in-plane, the single particle level ordering is reversed within the $t_{2g}$ and $e_{g}$ manifolds as shown in Fig.~\ref{fig1}(a) ({\em right panel}), with respect to the earlier situation. In terms of absolute values, the effective, distortion induced splittings are $\Delta_{e_{g}}$ $\sim$ 0.231 eV and $\Delta_{t_{2g}}$ $\sim$ 0.067 eV. As seen readily from Fig.~\ref{fig2}(b), this also in some sense reverses the nature of the many-body excited states with respect to the case of NiO on Ag. The first excited state involves the {\em spin-preserving} excitation of an electron from the $d_{xy}$ orbital to the $d_{x^2-y^2}$ orbital, which is connected to the ground state only via $L_{z}$. The second excited state, on the other hand, is a result of again a {\em spin-preserving} excitation, from the
the doubly degenerate $d_{xz},d_{yz}$ orbitals, jointly to the
 $d_{x^2-y^2}$ (60\%) as well as the higher lying $d_{3z^2-r^2}$ orbital (40\%), and this is connected to the ground state only via the $L_{x,y}$ operators. An analysis similar to the {\em contracted} case yields $\Lambda_{x}<\Lambda_{z}$, implying $D<0$, as required.
 
Returning to a similar analysis within a single-particle picture or that with realistic scalar $(U,J)$, we find for other parameters used in our calculations, that for the {\em contracted} case (Fig.~\ref{fig2}(a) ({\em left panel})), the {\em first and second excited states} correspond to the transitions $d_{3z^{2}-r^{2}}\rightarrow d_{x^{2}-y^{2}}$, {\em with a spin-flip}, and a {\em spin-preserving} $d_{xy}\rightarrow d_{3z^{2}-r^{2}}$ from the ground state, for both of which the matrix elements of all of $L_{x,y,z}$ vanish identically. So the first contributions arise from the {\em third} and {\em fourth} excited states which correspond to the spin-preserving transitions $(d_{xz}$, $d_{yz})\rightarrow d_{3z^2-r^2}$ and $d_{xy}\rightarrow d_{x^2-y^2}$, very similar to the {\em first and second} excited states for the same case with multiplets, as described above. Thus, similar to the real situation with multiplets, we would get $D>0$ (spin orients {\em in-plane}), though with a {\em reduced value of anisotropy}. On the other hand, for the {\em expanded} case, as shown in Fig.~\ref{fig2}(a) ({\em right panel}), the {\em first excited state} involves the {\em spin-flip} transition $d_{x^{2}-y^{2}}\rightarrow d_{3z^{2}-r^{2}}$ (matrix elements of all of $L_{x,y,z}$ vanish), while the {\em second excited state} corresponds to the {\em spin-preserving} transition $(d_{xz}$, $d_{yz})\rightarrow d_{x^{2}-y^{2}}$ (only matrix element of $L_{x,y}$ finite), implying $D>0$ and hence a spin-orientation {\em in-plane}. So this case is {\em qualitatively different} from the {\em real} situation with multiplets, and the above implies that there is no SRT possible in a theory which includes only anisotropic hopping and scalar $(U,J)$. 
Thus the use of a scalar $(U,J)$ gives us very different excited states 
compared to those obtained when we have the full multiplet interactions. 

\begin{figure}
\begin{center}
\includegraphics[angle=0, width=0.5\textwidth]{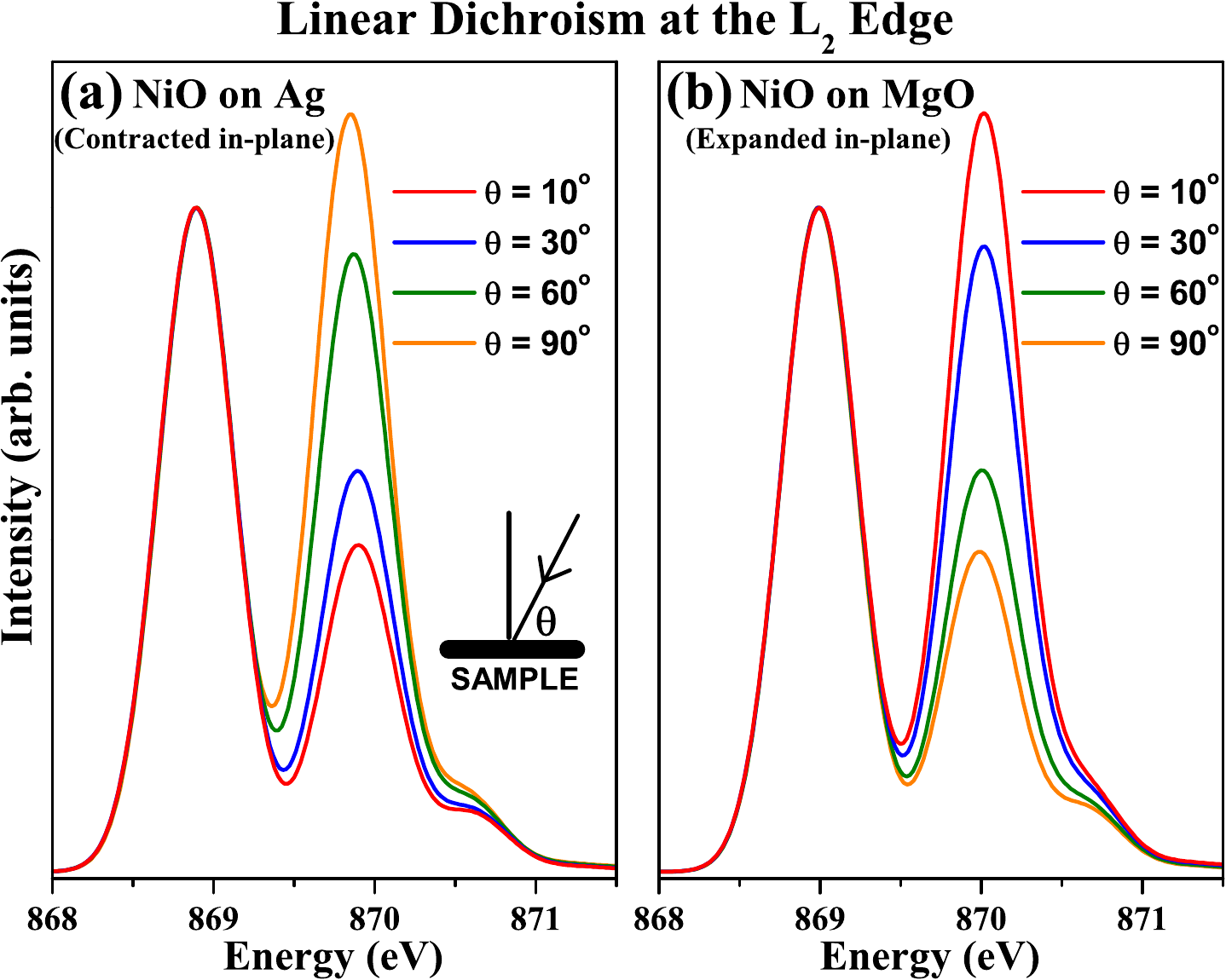}
\caption {(\emph{color online}) Calculated XMLD spectral trends as a function of the polar angle $\theta$ (and hence of the light polarization with respect to the surface normal), for the two-peaked structure at the $L_{2}$-edge (a) for the case of NiO on Ag ({\em contracted in-plane}), and (b) for the case of NiO on MgO ({\em expanded in-plane}). Clearly the trends are opposite in the two cases and reproduce experimental trends~\cite{NiO-MgO-expt,NiO-Ag-expt} well.} \label{fig3}
\end{center}
\end{figure}

As a final test of the model, we apply it, with suitable modifications~\cite{calc-methodology}, 
to calculate the trends in the x-ray magnetic linear dichroism (XMLD) for both the cases of NiO grown on MgO and on Ag substrates, for which experimental data 
is available~\cite{NiO-MgO-expt,NiO-Ag-expt}. Such calculations have been successful in describing the spectral trends also for the XMLD in
strained CoO thin films~\cite{xmld-spectra}.
The $L_{2}$ region (states excited with the $j=1/2$ core-hole) of the XAS for the Ni$^{2+}$
($d^{8}$) ion shows a distinct two peaked structure~\cite{NiO-MgO-expt}, and there is systematic transfer of spectral weight between the two features as the polarization of the incident light is changed from LP in the plane of the sample ($\theta$ $=$ 90$^{o}$) to LP along
the normal to the sample surface ($\theta$ $=$ 0$^{o}$) ($\theta$ is the angle between the direction of incident light and the sample surface, as shown in the 
{\em inset} to Fig.~\ref{fig3}(a)). If we denote by ${\cal R}$ the ratio of the peak intensities of the {\em first} with respect to the {\em second} $L_{2}$ peak, then it has been demonstrated that for NiO on Ag~\cite{NiO-Ag-expt}, ${\cal R}$ {\em decreases} with $\theta$, while for NiO on MgO~\cite{NiO-MgO-expt} ${\cal R}$ {\em increases} with $\theta$, as successfully reproduced in Fig.~\ref{fig3} (a) and (b), respectively.

In conclusion, we have set up a model that captures the substrate strain-driven SRT in NiO. While it is not surprising that strong Coulomb interactions modify the excitation spectra from what one would expect on single particle theory grounds, orbitally resolved multipole Coulomb interactions are found to determine the sign of the anisotropy constants and therefore the SRT in thin films of NiO. Thus, for thin films of correlated insulators, hopping anisotropy alone is not enough to determine the magnetocrystalline anisotropy and the observed SRT, but orbitally resolved anisotropic Coulomb interactions play a key role. The model also successfully reproduces XMLD spectral trends observed in experiments as a function of the sign and magnitude of substrate strain. 

We thank Prof. Krishnakumar S.R. Menon (SINP, India) for useful discussions regarding the experimental scenario. SSG would also like to acknowledge enlightening discussions with Prof. George A. Sawatzky (UBC, Canada). PM gratefully acknowledges the support provided by the Science and Engineering Research Board (SERB), under Project No. SERBPOWER (SPF/2021/000066). S.P. would like to express sincere thanks for the Ph.D. fellowship from the DST-INSPIRE, Government of India (IF190524).


\section{Supplementary Material for : {\em ``Spin orientation - a subtle interplay between strain and multipole Coulomb interactions"}}

\subsection{Many-body CI ground state and spectral calculations}

The valence Ni 3$d$ and O 2$p$ orbitals were included
for the (NiO$_6$)$^{10-}$ cluster that we considered ground state and spectral (XAS/XMLD) calculations, corresponding to the transition $2p^{6}3d^{8}\rightarrow 2p^{5}3d^{9}$. In addition, the core Ni 2$p$ orbitals were specially included for the spectral calculations that involve a core-hole in final state. The basis then consists of Slater determinants each corresponding to a particular way of distributing the two valence holes of Ni$^{2+}$ amongst the 46 valence spin-orbitals of Ni $3d$ and O $2p$, for the ground state calculations. For the spectral calculations, the basis involves distributing the single valence hole amongst the 46 valence spin-orbitals and the core-hole amongst the 6 core spin-orbitals. Each such basis state corresponds to an {\em electronic configuration}.

The Hamiltonian that was set up to describe the strained NiO films included appropriately scaled
hoppings between the valence orbitals (scaled according to Harrison's scaling laws~\cite{Harrison} with reference to the original hoppings for the undistorted structure), full multiplet (orbital and spin resolved) Coulomb interactions, $U_{\alpha\beta\gamma\delta}$, on the Ni 3$d$ orbitals as well as between the Ni 3$d$ and Ni 2$p$
orbitals. For $d$ electrons {\em all} such Coulomb integrals can be expressed in terms on only three Slater-Condon parameters $(F^{0}_{dd},F^{2}_{dd},F^{4}_{dd})$ with the {\em multiplet averaged} value being given by $U_{dd}=F^{0}_{dd}-\frac{2}{63}(F^{2}_{dd}+F^{4}_{dd})$.
Spin-orbit interactions were also included on the Ni
orbitals, in addition to a bare crystal field splitting of the Ni 3$d$
orbitals. All parameters entering the Hamiltonian were taken from
earlier estimates which were needed to obtain a good fit to the experimental
x-ray photoemission spectra~\cite{maiti}. Only the crystal-field splitting was slightly enhanced
from the earlier value to get a good description of the XMLD spectra. 

In addition to the above terms, we also need to include the antiferromagnetic correlations between the Ni spins at different lattice sites below $T_{N}$, in some effective way, in order to obtain a good agreement with the polarization dependent XAS spectral trends. As is usual practice, this is included in a single-site calculation, by means of a mean exchange field, pointing always in the direction of the sample magnetization, which is already decided by the inherent magneto-crystalline anisotropy in the system, and is hence not essential in understanding of the SRT. The exchange field couples to the spin degrees of freedom of the valence electrons only, as opposed to a {\em real} magnetic (Zeeman) field that couples to both spin and orbital degrees of freedom of {\em all} electrons. Full {\em configuration interaction} (CI) was carried out for the ground state as well as the excited state spectra. Calculations were carried out both {\em with} and {\em without} spin-orbit interactions on the Ni $3d$ orbitals. 

The latter were used as the starting point of the second-order perturbation treatment of
spin-orbit interaction, for the calculation of the anisotropy constant $D$, based on the equation :
\begin{equation}
\Lambda_{\mu\nu} = \sum_{n} \frac{\langle 0|L_{\mu}|n\rangle \langle n|L_{\nu}|0\rangle }{E_{n}-E_{0}},
~~(\mu,\nu = x,y,z), \label{lambda-tensor}
\end{equation} 
as discussed already in the text.
For this the matrix elements of $L_{\mu}$ ($\mu=x,y,z$) were calculated based on the exact eigenstates obtained from the cluster calculation, and the $\Lambda_{\mu\nu}$ tensor was evaluated based on these matrix elements as well as the many-body eigen-energies based on Eqn.~(\ref{lambda-tensor}), which was subsequently diagonalized to yield the anisotropy constant in terms of the principal components, as discussed in the paper. The perturbation parameter $\lambda$ used in the main article is related to the single electron spin-orbit parameter $\xi_{3d}$ for Ni, by the relation $\lambda=\frac{\xi_{3d}}{2S}$, where $S=1$ is the spin of the many-body ground state.

\subsection{Ab-initio results for Magnetocrystalline Anisotropy and the Spin-reorientation Transition}

The Ni-O equilibrium bondlengths needed for the above calculations,
were determined from {\it ab-initio} calculations for epitaxial
films. The in-plane lattice constant was fixed to that of the
substrate, while the out-of-plane lattice constant was obtained by
minimizing the total energy within WIEN2k \cite{wien2k}. 
The $xy$-plane of the NiO$_6$ cluster was chosen parallel to the substrate. The
generalised gradient approximation (GGA) with 
Perdew-Burke-Ernzerhof (PBE) form \cite{pbe}
of the functional was used for the exchange and a Monkhorst Pack
$k$-points grid of $10\times10\times10$ was used in the total energy minimization.

To ensure the accuracy of our calculations, convergence tests were performed with respect to both the plane-wave cutoff energy and the $k$-point mesh density. Initially, the cutoff energy was varied while keeping the $k$-point mesh fixed at $12\times12\times12$. Subsequently, with the cutoff energy set to $800~eV$, the $k$-point mesh was varied. Based on the convergence results, a cutoff energy of $800~eV$ and a $16\times16\times16$ $k$-point mesh were chosen for all subsequent calculations. The convergence data are summarized in the table below.
\begin{table}[h!]
\centering
\begin{tabular}{|c|c|}
\hline
Cutoff Energy Range (eV) & Change in $E$ (meV) \\
\hline
600 -- 700 & 15.75 \\
700 -- 750 & 2.21 \\
750 -- 800 & 1.17 \\
800 -- 850 & 0.0 \\
\hline
\end{tabular}

\vspace{0.5cm}

\begin{tabular}{|c|c|}
\hline
K-mesh & Change in $E$ (meV) \\
\hline
11 -- 12 & 20.84 \\
12 -- 13 & 9.12 \\
13 -- 14 & 1.48 \\
14 -- 15 & 1.54 \\
15 -- 16 & 0.01 \\
\hline
\end{tabular}
\caption{Convergence of total energy with respect to cutoff energy and k-point mesh.}
\end{table}
The magnetocrystalline anisotropy of NiO grown on MgO and Ag substrates was calculated both without and with the inclusion of Hubbard $U$ values of $2~eV$ and $6~eV$. In the absence of Hubbard $U$ ($U = 0$), a small gap opening near the Fermi energy was observed; however, the calculations failed to reproduce the correct spin reorientation behaviour. In contrast, incorporating $U = 2~eV$ and $6~eV$ successfully captured the experimentally observed spin reorientation trends of NiO across the different substrates.
\begin{table}[htbp]
\centering

\begin{tabular}{|c|c|c|c|c|}
\hline
Spin & NiO on MgO  &  NiO on Ag  & $\Delta E$ MgO  & $\Delta E$ Ag \\
direction & (in eV) & (in eV) & (in meV) & (in meV)\\
\hline
\multicolumn{5}{|c|}{\textbf{U=0}} \\
\hline
X & -45.634934 & -45.578936 & 0.09 & -0.09 \\
Y & -45.634934 & -45.578936 &      &      \\
Z & -45.634836 & -45.579028 &      &      \\
\hline
\multicolumn{5}{|c|}{\textbf{U=2}} \\
\hline
X & -42.393917 & -42.234290 & -0.34 & 0.69 \\
Y & -42.393917 & -42.234290 &      &       \\
Z & -42.394256 & -42.233605 &      &       \\
\hline
\multicolumn{5}{|c|}{\textbf{U=6}} \\
\hline
X & -37.844176 & -37.676759 & -0.36 & 0.71 \\
Y & -37.844176 & -37.676759 &      &       \\
Z & -37.844541 & -37.676051 &      &       \\
\hline
\end{tabular}
\caption{Magneto crystalline anisotropy for NiO on MgO and Ag substrates at different values of Hubbard $U$ applied to Ni 3$d$ obital.}
\end{table}

\subsection{Symmetry Analysis}

\begin{figure}
\begin{center}
\includegraphics[angle=0, width=0.5\textwidth]{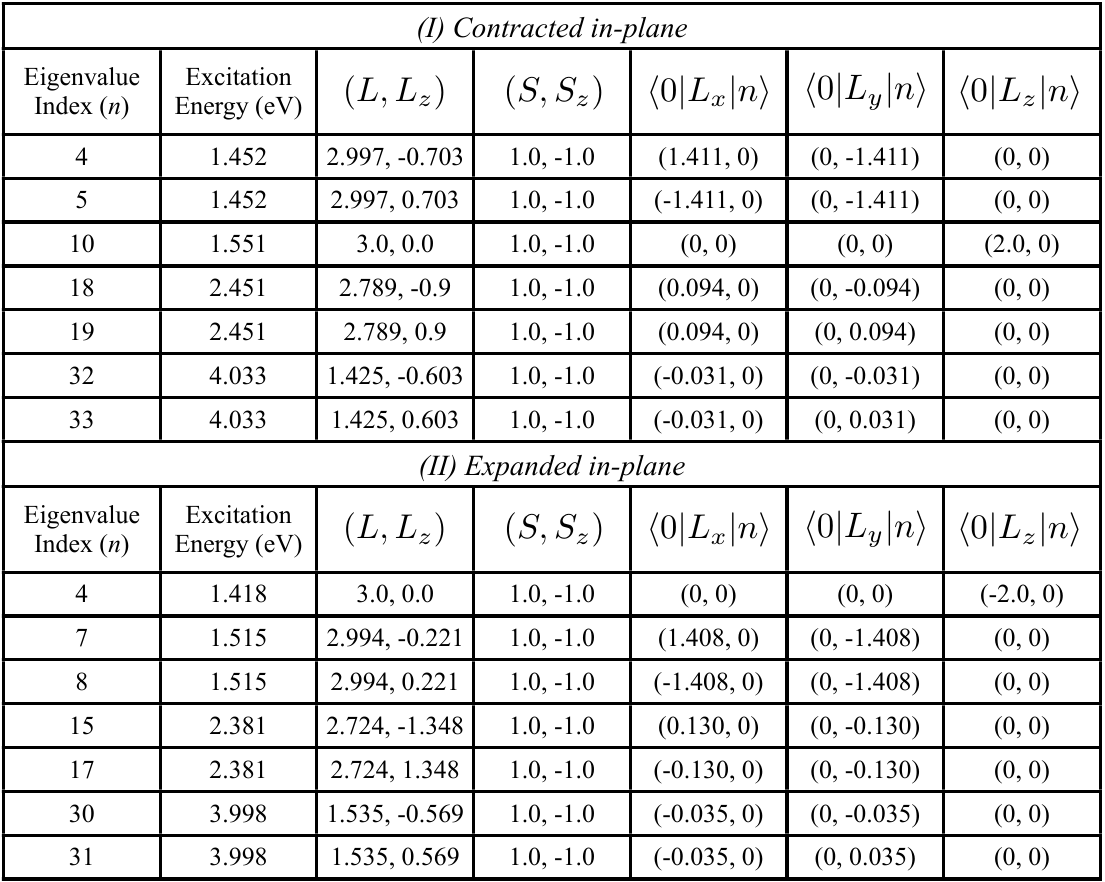}
\caption {Table of the excited states, $|n\rangle$, connected via {\em finite} matrix elements of the $L_{\mu}~(\mu=x,y,z)$ operators to the ground state, $|0\rangle$. The excitation energies ($E_{n}-E_{0}$), $(L,L_{z};S,S_{z})$ for these states, and the matrix elements $\langle 0|L_{\mu}|n\rangle$, are listed for the (I) {\em contracted in-plane}, and (II) for the {\em expanded in-plane} cases. In either case $(L,L_{z};S,S_{z})_{gnd}=(3,0;1,-1)$.} \label{table-matel}
\end{center}
\end{figure}

The finite matrix elements of the $L_{x,y,z}$ operator between the excited states and the orbitally quenched ground state, as described in the main article, along with their excitation energies and spin and orbital quantum numbers $(L,L_{z};S,S_{z})$, are shown in Fig.~\ref{table-matel}. We find that some of these states are doubly degenerate, differing only in their $\langle L_{z}\rangle$ values. Though the summation in Eqn.~(\ref{lambda-tensor}) extends over all higher excited states, one would generally expect contributions from higher excited states to be smaller by virtue of the energy denominator becoming larger. Interestingly, we find that it is only the first two excited states which determine the sign and magnitude of the anisotropy constants, while most of the higher excited states do not contribute, as the matrix elements of $L_{\mu}$ (the numerators in Eqn.~(\ref{lambda-tensor})) go to zero. The very few finite matrix elements which come from some of the higher excited states are at least an order of magnitude smaller. Those states for which the effective $L$ value deviates noticeably from the ground state value of $L=3$ are found to have smaller matrix elements, the larger this deviation, the smaller is the value of the {\em spin-orbit allowed} matrix elements. We will base our arguments on an atomistic picture with an ionic crystal field. This has essentially the same symmetries as the NiO$_{6}$ octahedron, but fewer states to consider. Here we describe the symmetry reasons for the very few spin-orbit allowed matrix elements.

The eigenstates of the $d^{8}$ configuration for Ni$^{2+}$, in spherical ($O_{3}$) symmetry, {\it i.e.,} in the presence of only multiplet Coulomb interactions consists of the terms $^{3}F$, $^{1}D$, $^{3}P$, $^{1}G$, and $^{1}S$, in that order in energy, so that the ground state is $^{3}F$. In the presence of crystal fields, as one descends in symmetry from $O_{3}$ to $O_{h}$ ({\em octahedral}) and finally to $D_{4h}$ ({\em tetragonal}), which is relevant here, group theory tells us that this term branches down to the following irreducible representations (IRREPS) : $$^{3}F \rightarrow~^{3}A_{2g} ~\oplus~ ^{3}T_{1g}~\oplus~ ^{3}T_{2g} \rightarrow~^{3}B_{1g} ~\oplus~ ^{3}A_{2g} ~\oplus~ 2\times^{3}E_{g} ~\oplus~ ^{3}B_{2g}$$ 
The energy ordering of these levels change as one goes from the contracted in-plane to the expanded in-plane case. 
$^{3}B_{1g}$ however remains the ground state in either case.

The higher lying Coulomb terms similarly branch down into various irreducible representations. This independent branching of the various Coulomb terms 
is however valid only for a {\em very weak} crystal field. On the other hand, in cases where the crystal field is sizeable ({\it e.g.,} in our case 
the effective $10Dq$ is $\sim$ 1.5 eV), it can cause configuration mixing between crystal field multiplets with the same value of total spin, $S$. The only 
other triplet ($S=1$) state is $^{3}P$ which in turn branches as follows : $$^{3}P \rightarrow~^{3}T_{1g} \rightarrow~^{3}E_{g} ~\oplus~ ^{3}A_{2g}$$ 
The crystal field, being a part of the Hamiltonian, can only mix terms with the same symmetry. This results in configuration interaction (CI) between 
the two $^{3}A_{2g}$ IRREPS from the $^{3}F$ and $^{3}P$, as well as between the two $^{3}E_{g}$ IRREPS of $^{3}F$ and the one from the $^{3}P$. 

For the calculation of $D$, based on Eqn.~(\ref{lambda-tensor}), we note that any states originating from the $^{3}P$ term ($L=1$) are not connected to the $^{3}B_{1g}$ ($L=3$) ground state normally via spin-orbit interaction, but for the aforesaid CI. Explicit calculations using lattice harmonics~\cite{balhausen} show that the {\em bare} transition $^{3}A_{2g}\rightarrow~^{3}T_{2g}$ is allowed via the spin-orbit term in $O_{h}$ symmetry, which branches down into the two {\em bare} transitions $^{3}B_{1g}\rightarrow~^{3}B_{2g}$ [$\langle ^{3}B_{1g}|L_{z}|^{3}B_{2g}\rangle=2$] and $^{3}B_{1g}\rightarrow~^{3}E_{g}$ [$\langle ^{3}B_{1g}|L_{x},L_{y}|^{3}E_{g}\rangle=\pm\sqrt{2},-\sqrt{2}i$], in $D_{4h}$ symmetry. These are the finite matrix elements listed in Fig.~\ref{table-matel}. We however notice that the transition amplitude is slightly reduced from $\sqrt{2}$ for the $^{3}E_{g}$ state, due to the aforesaid CI with the two other {\em bare} $^{3}E_{g}$ states originating from the {\em forbidden} $^{3}T_{1g}$ states (in $O_{h}$), one from the $^{3}F$ and the other from the $^{3}P$ term. The same CI also gives rise to the two {\em mildly allowed} transitions listed at higher energies in Fig.~\ref{table-matel}. Such symmetry analysis shows that only very few terms in the sum in Eqn.~(\ref{lambda-tensor}) contribute, in general, to the calculation of $D$, and that multipole Coulomb interactions play a pivotal role.

While such general systematics of matrix elements can be understood on the basis of pure symmetry arguments, the actual calculation of $D$ requires also the knowledge of the ordering of levels and the energies of the eigenstates in the two cases. This, as also the actual values of the matrix elements, in a situation that also treats covalency on an equal footing, are carried out much easily using the exact diagonalization scheme outlined before, and the results are described in the main article.

\end{document}